\documentclass[journal=ancac3,manuscript=article]{achemso}
\usepackage{graphicx}
\usepackage{amsfonts}
\usepackage{amsmath}
\usepackage{amssymb}
\usepackage{exscale}
\usepackage{afterpage}				
\usepackage{lineno}
\usepackage{xcolor}
\usepackage{natbib}
\usepackage{chemformula} 
\usepackage[T1]{fontenc} 
\usepackage{diagbox}

\title{Nanoscale characterization of atomic positions in orthorhombic perovskite thin films}

\author{M. Martirosyan}
\affiliation{Institut Jean Lamour, CNRS, Universit\'e de Lorraine, 54000 Nancy, France}
\altaffiliation{These authors contributed equally to this work.}

\author{S. Passuti}
\affiliation{CRISMAT, Normandie Universit\'e, CNRS, ENSICAEN, UNICAEN, 14000 Caen, France}
\altaffiliation{These authors contributed equally to this work.}

\author{G. Masset}
\affiliation{Institut Jean Lamour, CNRS, Universit\'e de Lorraine, 54000 Nancy, France}

\author{J. Varignon}
\affiliation{CRISMAT, Normandie Universit\'e, CNRS, ENSICAEN, UNICAEN, 14000 Caen, France}

\author{H. Chintakindi}
\affiliation{Department of Structure Analysis, Institute of Physics of the Czech Academy of Sciences, Na Slovance 2, 18200 Prague, Czechia}

\author{J. Ghanbaja}
\affiliation{Institut Jean Lamour, CNRS, Universit\'e de Lorraine, 54000 Nancy, France}

\author{S. Migot}
\affiliation{Institut Jean Lamour, CNRS, Universit\'e de Lorraine, 54000 Nancy, France}

\author{A. Benedit-C\'ardenas}
\affiliation{Institut Jean Lamour, CNRS, Universit\'e de Lorraine, 54000 Nancy, France}

\author{L. Pasquier}
\affiliation{Institut Jean Lamour, CNRS, Universit\'e de Lorraine, 54000 Nancy, France}

\author{K. Dumesnil}
\affiliation{Institut Jean Lamour, CNRS, Universit\'e de Lorraine, 54000 Nancy, France}

\author{L. Palatinus}
\affiliation{Department of Structure Analysis, Institute of Physics of the Czech Academy of Sciences, Na Slovance 2, 18200 Prague, Czechia}

\author{W. Prellier}
\affiliation{CRISMAT, Normandie Universit\'e, CNRS, ENSICAEN, UNICAEN, 14000 Caen, France}

\author{A. David}
\affiliation{CRISMAT, Normandie Universit\'e, CNRS, ENSICAEN, UNICAEN, 14000 Caen, France}

\author{Ph. Boullay}
\email{philippe.boullay@ensicaen.fr}
\affiliation{CRISMAT, Normandie Universit\'e, CNRS, ENSICAEN, UNICAEN, 14000 Caen, France}

\author{O. Copie}
\email{olivier.copie@univ-lorraine.fr}
\affiliation{Institut Jean Lamour, CNRS, Universit\'e de Lorraine, 54000 Nancy, France}

\begin{document}

\newpage

\begin{abstract}
The crystal structure determines many of the physical properties of oxide perovskites (ABO$_3$) and only a tiny modification of the lattice structure causes major changes in the functional properties through the interplay among spin, orbital and charge orders. The determination of characteristic distortions and symmetries is a valuable asset for understanding the structure-properties relationship and guiding the design of epitaxial oxide heterostructures, where electron degrees of freedom and correlated electronic states can be tailored. Even until new phases, otherwise absent in bulk materials, may appear. Here, we report on the in-depth structural characterization of 50~nm-LaVO$_3$ thin film grown onto (110)-oriented DyScO$_3$ by molecular beam epitaxy. We have investigated the heterostructure by means of x-ray diffraction, high-resolution and scanning transmission electron microscopies, scanning precession electron diffraction tomography and first-principle calculations. LaVO$_3$ crystallizes in the orthorhombic $Pbnm$ space group and is constrained by the substrate, which imposes a growth along the $[110]$ orthorhombic direction, over the 140 deposited unit cells. The mapping of the reciprocal space allows determining the orientation of the film and refining the lattice parameters. Using scanning transmission electron microscopy, we analyzed the structure of LaVO$_3$, focusing on the determination of the antipolar displacement of the rare earth. Additionally, 3D electron diffraction enabled to resolve the atomic positions of all species within the film. 
\end{abstract}
%
%
%
%
\maketitle
%
%
%
%
%

\section{INTRODUCTION}

Beyond the undistorted $Pm\bar{3}m$ cubic structure, the compounds with a perovskite structure can crystallize in various space groups. In particular, transition metal oxides perovskites with the chemical formula ABO$_3$ often crystallize in an orthorhombic structure owing to the ability of the perovskite structure to distort and accommodate different A-site species (alkali, alkaline earth and rare earth metals) and B-site site species (transition metals), both with different atomic radii. To accommodate the mismatch between A--O and B--O bond lengths, many transition metal ABO$_3$ perovskites crystallize in the $Pbnm$ space group, characterized by cooperative oxygen octahedra rotations. Due to the change in the A--O covalency, the rotations of the BO$_6$ octahedra lead to a shift of A cations from their position in cubic symmetry~\cite{Goodenough1971}. All these atomic displacements resulting from structural changes contribute to the interplay between the charge, orbital and spin degrees of freedom of the electrons and give rise to complex orbital and spin ordered phases~\cite{Imada1998,Tokura2000}. In the $Pbnm$ space group, the cooperative BO$_6$ octahedra rotations are described as $a^-a^-c^+$ tilt pattern according to Glazer's notation~\cite{Glazer1972}: \textit{i.e.} out-of-phase rotations along the $[110]$ direction and in-phase rotation along the $[001]$ direction. 

Tuning the bond angles and lengths in epitaxially strained heterostructures is an efficient way for tailoring the functional properties of ABO$_3$ systems and even promoting new properties otherwise absent in the bulk material~\cite{Haeni2004}. These include the RVO$_3$ series (R is a rare earth or yttrium). Indeed, the spins of the V$^{3+}$ $3d$ electrons, occupying the $t_{2g}$ orbitals, can order at a temperature higher than the bulk material, when PrVO$_3$ is compressively strained on LaAlO$_3$ ($-2.9\%$ lattice mismatch)~\cite{Kumar2019a, Kumar2019b}. When grown on SrTiO$_3$ ($+0.1\%$ lattice mismatch)~\cite{Copie2013}, the tuning of the intra-plane and inter-plane magnetic superexchange interactions between nearest neighbors V$^{3+}$ yields a non-monotonous evolution of spin ordering temperature through uniaxial strain engineering~\cite{Copie2017}. The epitaxial strain has also an effect on the orbital transition, as in LaVO$_3$ thin films~\cite{Meley2021} and YVO$_3$/LaAlO$_3$ superlattices~\cite{Radhakrishnan2021,Radhakrishnan2022}. RVO$_3$ is a prototypical platform for investigating $t_{2g}$ orbital physics, interactions with spin orders, and the couplings with structural distortions, such as relatively weak Jahn-Teller (JT) distortions~\cite{Miyasaka2003}. This JT distortion, consisting essentially in long and short V$-$O bonds length along the pseudo-cubic directions~\cite{Zhou2005}, allows lifting the degeneracy between the V $t_{2g}$ levels. Hence, the two electrons per V$^{3+}$ site are located either in the $d_{xz}$ or in the $d_{yz}$ orbitals, $d_{xy}$ being always occupied. As the temperature decreases, RVO$_3$ undergo a structural phase transition from orthorhombic $Pbnm$ to monoclinic $P2_1/b$~\cite{Sage2007,Miyasaka2003}. Concurrently, the cooperative JT distortions promote a G-type orbital ordering ($\rm G_{OO}$). At low temperature, superexchange interactions stabilize then a C-type antiferromagnetic spin ordering of the V spins ($\rm C_{SO}$). Since controlling the properties of RVO$_3$ is important for their potential applications, it is essential to delve into the structure of this compound deposited as thin films and to obtain reliable atomic positions for properties modeling or artificial intelligence data processing. 

Here, we report on the epitaxial growth of LaVO$_3$ (LVO) by molecular beam epitaxy on (110)-oriented DyScO$_3$ (DSO) substrates and the in-depth characterization of the film structure. In the literature, LVO thin films have been grown by pulsed laser deposition (PLD) on various substrates~\cite{Masuno2004, Hotta2006} and by MBE~\cite{Zhang2015, Masset2020}, with a strong interest in the electronic~\cite{Hotta2007,Zhou2007} and optical properties~\cite{Wang2015,Zhang2017}. The occurrence of hybrid improper ferroelectricity~\cite{Bousquet2008,Benedek2011,Rondinelli2012} in short-period (LaVO$_3$/PrVO$_3$) superlattices has been predicted with an electrical control of orbital orderings, natively coupled to the spin orders~\cite{Varignon2015}. The proposed ferroelectric character of the superlattice is due to the coupling between VO$_6$ rotations and R-cation in-plane anti-polar motions (X$_5^-$ mode) and between the JT distortions and the out-of-plane anti-polar motions (X$_3^-$ mode). In symmetry-breaking RVO$_3$ superlattices, X$_5^-$ and X$_3^-$ motions become polar P$_{xy}$ and P$_{z}$ motions, respectively. However, this way to design ferroelectricity in otherwise non-ferroelectric $Pbnm$ systems requires the $[001]$ orthorhombic direction to be along the growth direction~\cite{Varignon2015}. While compressively strained LVO is (110)-oriented and multidomain when grown on (001)-oriented SrTiO$_3$~\cite{Rotella2012,Rotella2015,Masset2020}, tensely strained LVO is (001)-oriented when grown on (110)-oriented DSO by PLD~\cite{Meley2018}. Recently, the design of spin and orbital ordering patterns in YVO$_3$ was achieved by growing 80 unit cells onto different surface orientations of an orthorhombic substrate, with specific octahedral rotations and cation displacements~\cite{Radhakrishnan2024}. However, the demonstration of a competition between film/substrate oxygen octahedral connectivity and epitaxial strain, induced by the lattice mismatch in LVO grown on (110)-oriented DSO, revealed that the film and substrate orientations can differ beyond a critical thickness of approximately 70 unit cells~\cite{Alexander2024}. 

Both LVO and DSO crystallize in the $Pbnm$ (space group no. 62). The lattice parameters are $a_o = 5.555$~\AA, $b_o = 5.549$~\AA, $c_o = 7.849$~\AA~for LVO~\cite{Bordet1993,Sage2007}, and $a_s = 5.438$~\AA, $b_s = 5.714$~\AA, $c_s = 7.897$~\AA~for DSO~\cite{Meley2018}. Based on the simple consideration of the pseudo-cubic lattice parameters of bulk DSO and LVO, we can calculate the biaxial strain $\epsilon$ experienced by the LVO film for different orientations, \textit{i.e.} the LVO $[001]$ direction is parallel or perpendicular to the growth direction. For the latter, the $[001]$ can be parallel or perpendicular to the DSO $[001]$ direction. Hence, for LVO$[110]\|$DSO$[110]$, we obtain $\epsilon_{001}=0.61\%~(0.54\%)$ and $\epsilon_{\bar{1}10}=0.43\%~(0.50\%)$ for the LVO $[001]$ direction parallel (perpendicular) to the in-plane DSO $[001]$ direction. For LVO$[001]\|$DSO$[110]$ we have $\epsilon_{\bar1{1}0}=0.54\%$ and $\epsilon_{110}=0.43\%$. Hence, these strain values leads us to anticipate the $[001]$ axis orientation parallel to the growth direction, as it should minimize the strain energy. Here, we however observe that a 50-nm thick LVO film is (110)-oriented.

\section{Results and discussion}
%
%
%
%
%
\begin{table*}[tp]
\caption{\label{Table1} Orientation axis optimizations of LVO on (110)-DSO from first-principles calculations and corresponding spin orders, space groups, amplitudes of the structural distortions, given in \AA~per formula unit, from the calculated structures. The last column shows the energy differences, given in meV per formula unit and compared to the first line.}
\begin{tabular}{ccccccccc}
\hline
axis & spin order & space group & $a^-a^-c^0$ & $a^0a^0c^+$ & X$_5^-$ & Q$_2^+$ & Q$_2^-$ & $\Delta\rm E$  \\
\hline
$[001]$ & ${\rm C_{SO}}$ & $P2_1/c$ & 0.678 & 0.393 & 0.183 & 0.014 & 0.042 & 0  \\
$[001]$ & ${\rm G_{SO}}$ & $Pbnm$ & 0.676 & 0.407 & 0.197 & 0.023 & 0.000 & -3  \\
$[110]$ & ${\rm G_{SO}}$ & $P2_1/m$ & 0.668 & 0.455 & 0.272 & 0.044 & 0.000 & -39  \\
$[110]$ & ${\rm C_{SO}}$ & $P\bar{1}$ & 0.654 & 0.454 & 0.254 & 0.003 & 0.045 & -40  \\
\hline
\end{tabular}
\end{table*} 

We have first investigated the LVO phases by optimizing, with first-principles simulations using density functional theory (DFT), the structure for two growth orientations and two possible magnetic orders (see Methods). In Table~\ref{Table1}, we show the results of the geometry relaxations considering different spin orders. The ground state corresponds to the $[110]$ directions of LVO and DSO being parallel. The corresponding magnetic ground state is $\rm C_{SO}$, in agreement with the literature~\cite{Miyasaka2003}. However, $\rm G_{SO}$ magnetic state is also possible, whose energy is only 1~meV/f.u. higher. These states are associated to different structure symmetries, $\rm C_{SO}$ with  $P\bar{1}$ and $\rm G_{SO}$ with $P2_1/m$, corresponding to $P2_1/b$ and $Pbnm$ for the bulk material~\cite{Varignon2015,Copie2017}, respectively. Although the $[110]$-oriented LVO structure is the most stable, the small energy separation between the $\rm C_{SO}$ and $\rm G_{SO}$ magnetic states indicates actually two metastable states at 0 K, which are associated to different crystal structure symmetries.
\begin{figure*}[tp]
    \begin{center}
       \includegraphics[trim={0cm 3cm 0cm 3cm}, clip,width=1\textwidth]{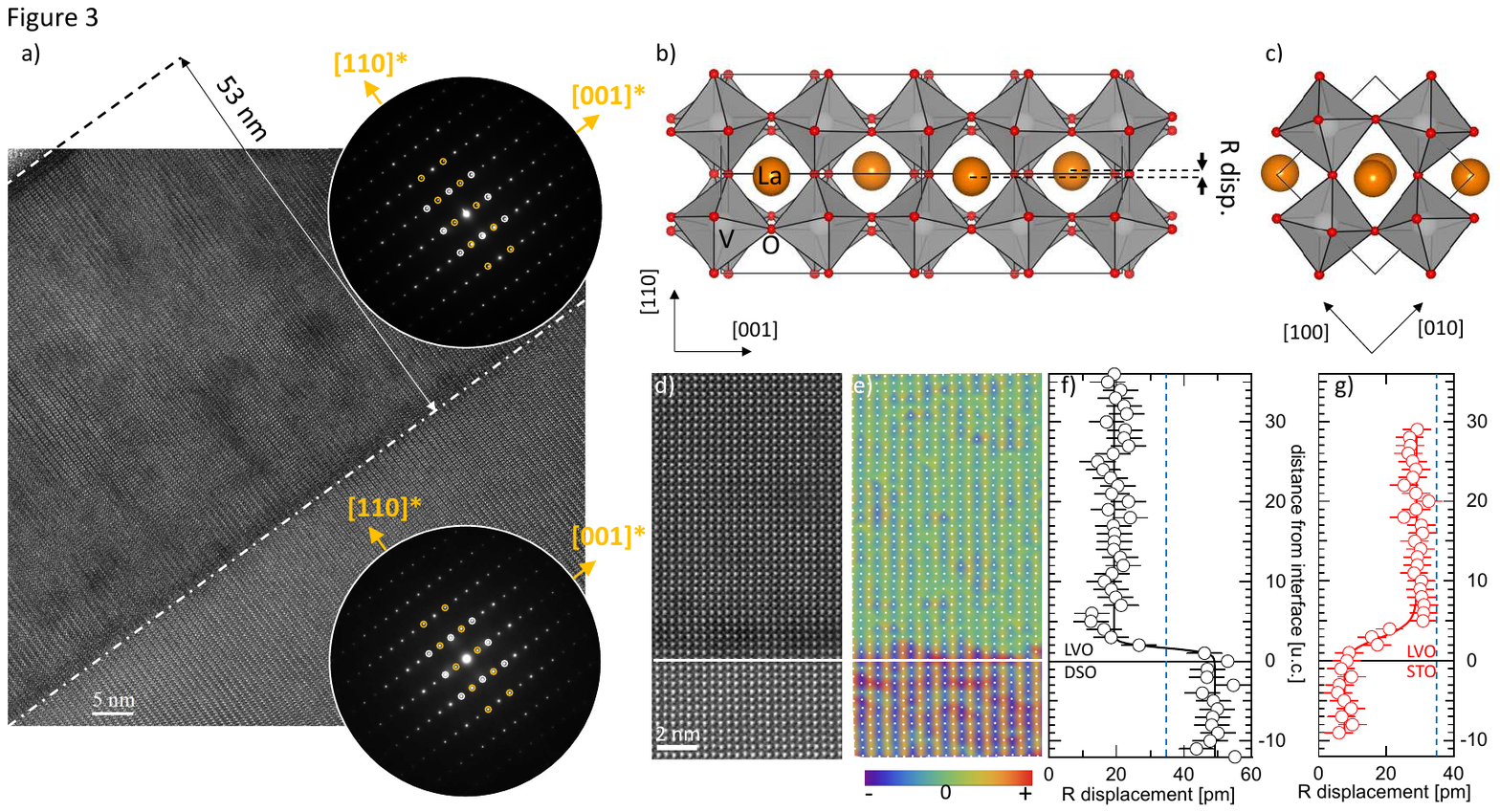}
    \end{center}
\caption{(a) HRTEM cross-section view along the $[\bar{1}10]$ DSO zone axis. EDP of the LVO and DSO region are displayed on the top and bottom of the image, respectively. (b) and (c) sketches of the LVO atomic structure projected along the $[\bar{1}10]$ and $[001]$ directions, respectively. (d) HAADF-STEM observation along the $[\bar{1}10]$ DSO zone axis. (e) mapping of the vertical La-cation displacements from HAADF-STEM image analysis. (f) and (g) mean R-site cation displacement across the growth direction for LVO grown on DSO and STO, respectively. The blue dashed line signals the displacement value in the bulk LVO (34 pm).}
\label{Fig1} 
\end{figure*}
We have also performed a symmetry mode analysis~\cite{Orobegoa2009,Perez-Mato2010} of the structures. The results are summarized in Table~\ref{Table1}. At 0 K, the derived structures are made of the oxygen octahedral rotation modes ($a^-a^-c^0$ and $a^0a^0c^+$ in Glazer's notation~\cite{Glazer1972}), the X$_5^-$ mode and the cooperative JT modes (Q$_2^+$ and Q$_2^-$). In-phase Q$_2^+$ mode favors a C-type ordering (C$\rm_{OO}$) of the V$^{3+}$ occupied orbitals, while out-of-phase Q$_2^-$ mode favors a G-type ordering (G$\rm_{OO}$). For the $[001]$ growth direction and in agreement with the K\"ugel-Khomskii mechanism~\cite{Kugel1973}, the Q$_2^+$ and the Q$_2^-$ dominate for the C$\rm_{SO}$ and G$\rm_{SO}$ spin orders, respectively. For the $[110]$ growth direction, the out-of-phase and in-phase JT modes dominate for the $P\bar1$ ($P2_1/m$ in bulk) and $P2_1/m$ ($Pbnm$ in bulk) structure symmetries, respectively. This agrees with the corresponding C$\rm_{SO}$/G$\rm_{OO}$ and G$\rm_{SO}$/C$\rm_{OO}$ spin/orbital states. In the case of $[110]$-oriented growth, we observe that the amplitude of the $a^-a^-c^0$ mode has decreased while those of the $a^0a^0c^+$ and X$_5^-$ have increased when compared to the $[001]$-oriented growth.

\begin{figure*}[tp]
    \begin{center}
    	   \includegraphics[trim={2.5cm 2cm 3cm 2cm}, clip,width=0.9\textwidth]{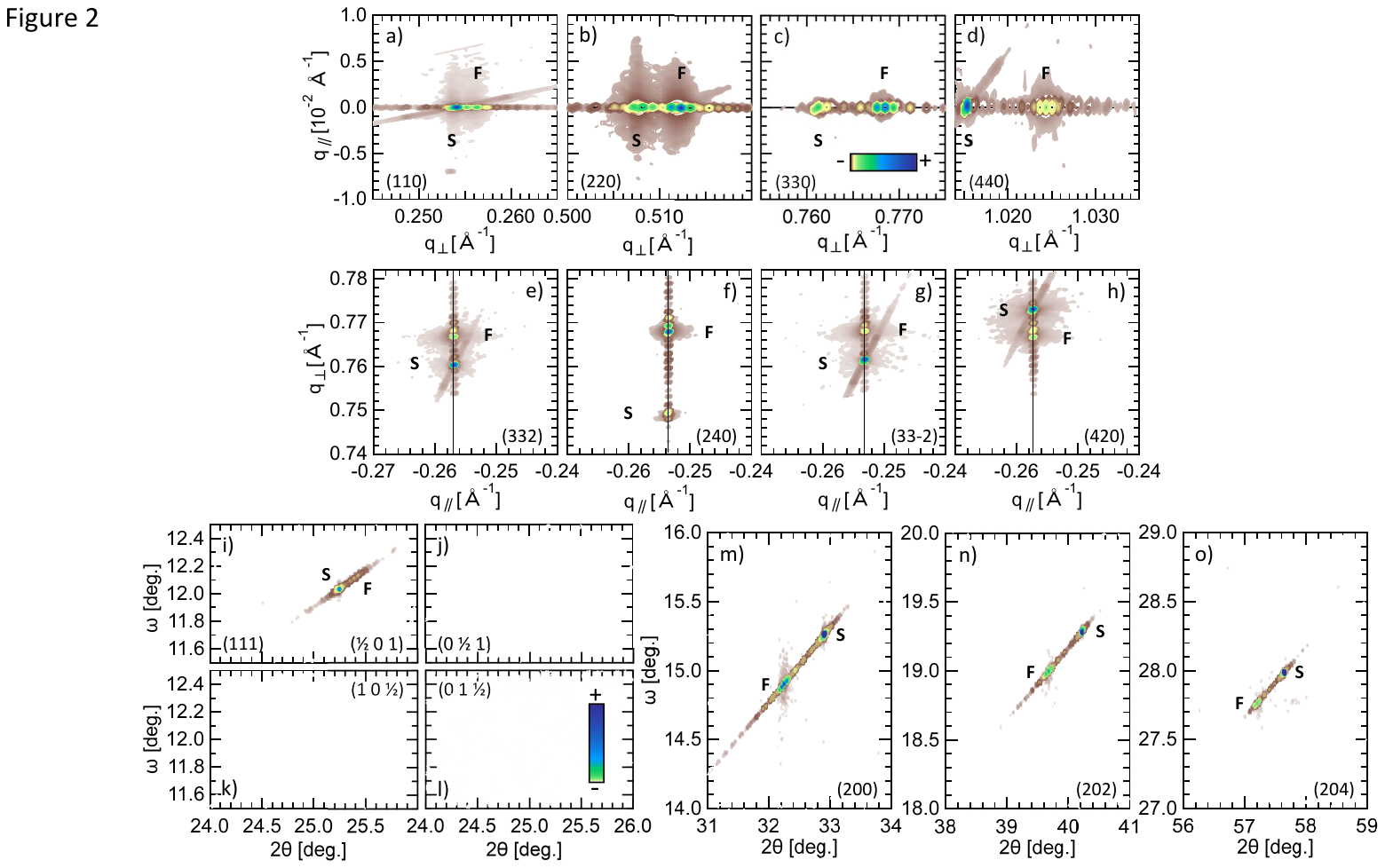}
    \end{center}
\caption{(a)-(b) Reciprocal space maps around the symmetric (110), (220), (330) and (440) Bragg peaks, respectively. (e)-(h) Reciprocal space maps around the asymmetric (332), (240), (33$\bar2$) and (420) Bragg peaks, respectively. (i)-(l) Reciprocal space maps around the (111) Bragg peak, or pseudo-cubic half-order Bragg peak $(\frac{1}{2}01)$, $(0\frac{1}{2}1)$, $(10\frac{1}{2})$ and $(01\frac{1}{2})$, respectively. (110)-oriented LVO displays only the $(\frac{1}{2}01)$ peak. (m)-(o) Reciprocal space maps around the (200), (202) and (204) Bragg peak. }
\label{Fig2} 
\end{figure*}

%
%
%
%
%
%
%
%
Guided by the theoretical investigation of a LVO film on DSO (we note that first principle calculations are performed below the structural phase transition temperature~\cite{Bordet1993}), we have then deposited 140 unit cells of LVO onto (110)-oriented DSO by MBE in a background pressure of distilled ozone (see Methods). A layer-by-layer growth mode, monitored by \textit{in situ} Reflective High Energy Electron Diffraction (RHEED), and a stoichiometric incorporation of oxygen during the growth indicated a high crystalline quality of the film (see Supplementary Information).

To get into the details of the growth of LVO on (110)-DSO, we have investigated the film's microstructure by High-Resolution Transmission Electron Microscopy (HRTEM). The observations were performed on the LVO cross-section taken along the $[1\bar{1}0]$ zone axis of DSO (see Methods). We show a representative HRTEM image in Fig.~\ref{Fig1}(a), where we observe a sharp interface (delimited by a white dotted and dashed line) between DSO (bottom) and LVO (top), whose film thickness is around 53~nm, \textit{i.e.} 136 unit cells, consistent with the RHEED and XRD experiments (see Supplementary Information). In Fig.~\ref{Fig1}(a), we show the electron diffraction patterns (EDP) of LVO and DSO, which exhibit reflections compatible with a $[1\bar{1}0]$ zone axis pattern. The reflections highlighted in orange, distinct from those of the cubic perovskite subcell (highlighted in white), indicate the presence of a supercell associated with a doubling of the cell parameter along the $[001]$ direction expected for the $Pbnm$ space group. In the bulk material, this doubling is induced by the $a^-a^-c^+$ octahedral tilt pattern and the displacements of the rare earth. The doubling of the cell parameter along the $[001]$ direction appears as a striped contrast perpendicular to the $[001]$ direction, in the HRTEM image, both in LVO and DSO. Hence, the LVO film follows the orientation imposed by DSO with the epitaxial relationship being LVO$[110]\|$DSO$[110]$ and LVO$[001]\|$DSO$[001]$. This observation agrees with our DFT calculations. In Fig.~\ref{Fig1}(b) and (c), we show a sketch of the bulk LVO atomic structure projected along the $[\bar{1}10]$ and $[001]$ directions, respectively. The characteristic antipolar motion of the La cations along the $[110]$ direction is observed, shifting alternatively upward and downward along the $[001]$ direction (X$_{5}^{-}$). For this projection, the apparent distance between consecutive shifted cations is represented by the parameter R$_{\rm{disp}}$. In Fig.~\ref{Fig1}(d), we show an image captured by Scanning Tranmission Electron Microscopy (STEM) in High-Angle Annular Dark-Field (HAADF) mode. As this technique is most sensitive to heavy elements, we can actually visualize and map the antipolar La motions~\cite{Masset2020}. In order to highlight the atomic positions, we carried out a numerical analysis of the raw STEM-HAADF image~\cite{Nord2017}. The atomic positions are extracted and compared with an undistorted reference, which is the average of the vertical and horizontal positions. 
The mapping of the La displacements and amplitudes on this projection allows the determination of both in-plane and out-of-plane octahedral tilts, \textit{i.e.} along the $[001]$ and $[110]$ directions, respectively. Focusing on the atomic displacement map in Fig.~\ref{Fig1}(e), we observe an alternating contrast in the $[001]$ direction and across the substrate/film interface. This characterizes the X$_{5}^{-}$ motion, which is present in both DSO and LVO, although the amplitudes are different. As the antipolar motion and oxygen octahedra rotation are coupled by a trilinear energetic term~\cite{Varignon2015}, this also shows the continuity of the oxygen octahedra rotation across the interface. In Fig.~\ref{Fig1}(f), we show the amplitude of R cation displacement (R$_{\rm{disp}}$) along the growth direction, as a function of the distance from the interface. Owing to its small ionic radius, Dy$^{3+}$ promotes high-angle oxygen octahedra rotations in DSO, which in turn favor large Dy displacements, as we see below the interface in Fig.~\ref{Fig1}(f). Crossing the interface, the amplitude of the La displacement drops down to 20~pm after a region of ca. 3 unit cells where the LVO film experiences the effect of the larger structural distortions of the substrate. We note that this differs from the DFT calculations. As a comparison, we display in Fig.~\ref{Fig1}(g) the case of LVO grown on (001)-oriented SrTiO$_3$ (STO) (see details in Ref.~\cite{Masset2020}), where no structural distortion exists at room temperature for STO. In this case, there should be no Sr atom displacement in the STO part. In Fig.~\ref{Fig1}(g), the displacement is not zero and gives then an estimation of the uncertainty which is close to standard deviations. Crossing the interface, the amplitude of the La displacement increases up to 30~pm over ca. 6 unit cells. The mapping of the R displacements and amplitudes, obtained from the $[\bar{1}10]$ $Pbnm$ orientation, allows the determination of the in-phase octahedral tilts, along the $[001]$ direction as well. Over a short length-scale close to film/substrate interface, the octahedra rotations gradually decreases or increases for DSO and STO, respectively. This shows the importance of the coherent connectivity of the oxygen octahedra at the interface that allows a octahedral rotational control~\cite{Rondinelli2012}, exploiting the octahedra tilt amplitude and pattern mismatches. Based on this initial finding, a clear difference is already evident with (110)- and (001)-oriented LVO phases separated by a switching plane~\cite{Alexander2024}. Driven by energetics, the transition to two-phased film occurs beyond a critical thickness between 60 and 74~unit cells deposited by PLD. However, we observe here a single (110)-oriented LVO film with a thickness up to about 140 unit cells deposited by MBE. This underscore the influence of the deposition technique employed which, as the kinetics of the deposited species, along with the growth temperature and pressure~\cite{Ramesh1990, Hamet1992}, collectively affect the growth mechanism.  

%
%
%
%
\begin{table}[tp]
\caption{\label{Table2} (left) Absolute $2\theta$ positions of the listed $(hkl)$ Bragg peaks. (right) LVO refined lattice constants.}
\begin{tabular}{cc|ccc}
\hline
$(hkl)$ & $2\theta$ (deg.)& lattice & ortho. & mono. \\
\hline
$(110)$ & 22.767(1) & param. & $Pbnm$ & $P2_1/m$ \\
$(111)$ & 25.425(1) & \\
$(200)$ & 32.247(1) & $a$ (\AA) & 5.527(5) & 5.548(2) \\
$(202)$ & 39.268(1) & $b$ (\AA) & 5.521(5) & 5.553(2) \\
$(220)$ & 46.485(1) & $c$ (\AA) & 7.908(4) & 7.891(2) \\
$(204)$ & 57.261(1) & $\alpha$ ($^{\circ}$) & 90 & 90 \\
$(330)$ & 72.591(1) & $\beta$ ($^{\circ}$) & 90 & 90 \\
$(240)$ & 77.083(1)& $\gamma$ ($^{\circ}$) & 90 & 90.59(8) \\
$(332)$ & 77.105(1) &$\epsilon_{110}$ ($\%$) & $-0.54$ & $-0.06$ \\
$(33\bar{2})$ & 77.105(1) & $\epsilon_{001}$ ($\%$) & $+0.75$ & $+0.54$ \\
$(420)$ & 77.126(1) \\
$(440)$ & 104.192(1) & \\
\hline
\end{tabular}
\end{table}

Having characterized the structure of LVO at the atomic scale, we performed high-resolution XRD (see Methods) to determine how epitaxial strains affect the LVO structure on a larger scale. We show in Fig.~\ref{Fig2}(a)-(d) a series of reciprocal space maps (RSM) measured around the symmetric LVO $(110)$, $(220)$, $(330)$ and $(440)$ Bragg peaks, respectively. Figures~\ref{Fig2}(e)-(h) display the asymmetric $(332)$, $(240)$, $(33\bar{2})$ and $(420)$ Bragg peaks, respectively. This series of Bragg peaks show that the LVO film (referred as \textbf{F}) was grown coherently on DSO (referred as \textbf{S}). The high crystalline quality of the film is attested by Laue fringes that are clearly seen for all film's peaks, whose width along $q_{||}$ is rather narrow. In an orthorhombic unit cell, the difference between the $a$- and $b$-lattice constants appears as a difference in $(240)$ and $(420)$ Bragg peak position along $q_{\perp}$, as for DSO in Fig.~\ref{Fig2}(f) and (h). However, the peaks of LVO show almost identical positions, within our experimental accuracy. This is in good agreement with a LVO film experiencing a tensile stress leading to $a\approx b$~\cite{Vailionis2011}. To get further insights, we have also measured additionnal Bragg peaks displayed in Fig.~\ref{Fig2}(i)-(o): $(111)$, $(200)$, $(202)$ and $(204)$, respectively. Assuming the $Pbnm$ space group, the $(111)$ peak corresponds to one of the half-order Bragg peak in the $Pm\bar{3}m$ space group: $(\frac{1}{2}01)$, $(0\frac{1}{2}1)$, $(10\frac{1}{2})$ and $(01\frac{1}{2})$. These peaks, originating from the quadrupled perovskite unit cell, depend on the orientation and the population of the domain present in the film. In the Fig.~\ref{Fig2}(i)-(j), we observe that only the $(\frac{1}{2}01)$ reflection exists, while the others are absent. This means the LVO film consists in a single orientation variant, determined by the (110)-oriented DSO substrate, which imposes its crystallographic orientation on a large scale. The structural coupling at the interface is facilitated by both LVO and DSO having the same $Pbnm$ space group with similar structural distortions, beyond the mere consideration of the strain. From the absolute $2\theta$ position of all Bragg peaks, we have refined the unit cell lattice parameters~\cite{celref}. The $2\theta$ position list and the unit cell parameters are summarized in Table~\ref{Table2}. We have first considered the $Pbnm$ space group and we find the $a$- and $b$-lattice parameters are shortened and almost equal. While, $\epsilon_{110}$ describes a compressive strain, in agreement with the experimental value deduced from XRD ($-0.61\%$, see Supplementary Information), $\epsilon_{001}$ has increased compared the expected value (see above). Thereby, considering the group-subgroup relationship~\cite{Howard1998}, we have also calculated the unit cell parameters assuming the $P2_1/m$ space group. This introduces an additional degree of freedom, allowing the angle between the free $a$- and $b$-axis to accommodate the strain, along with  the $a$ and $b$ axis lengths~\cite{Vailionis2011, Rotella2012}. Compared to the volume of the bulk material (242.1~\AA$^3$), it is equal to 241.3~\AA$^3$ and 243.1~\AA$^3$ for the $Pbnm$ and $P2_1/m$ space groups, respectively. In Table~\ref{Table2}, we find that the strain in the $[001]$ direction is closer to the expected value (see above), with $\gamma$ different but close to 90$^{\circ}$. But, the strain along the $[110]$ direction is now very different from the experimental value ($-0.61\%$, see Supplementary Information). As the monoclinic angle is very close to 90$^{\circ}$, it remains difficult to distinguish between the $P2_1/m$ and $Pbnm$ space groups, also due to the small number of accessible Bragg peaks. We have thus deepened further the structure analysis of the LVO thin film.

%
%
%
%
%
\begin{figure}[tp]
    \begin{center}
        \includegraphics[trim={5cm 5.0cm 5cm 3cm}, clip,width=0.45\textwidth]{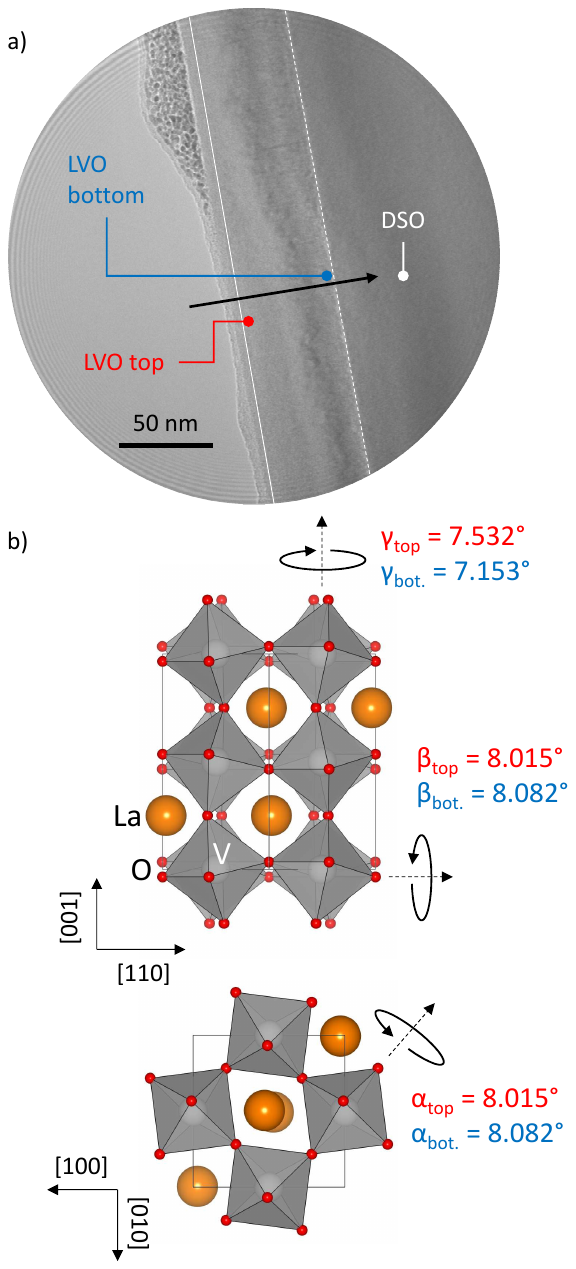}
    \end{center}
\caption{(a) STEM-ABF image of the scanned region along the black arrow and the refined area: LVO top, LVO bottom and DSO indicated in white. (b) Schematic representation of the structure of LVO as obtained from PEDT refinements, with the values of the octahedral tilt and rotations.}
\label{Fig3} 
\end{figure}

To accurately determine the atomic structure of the LVO film, structural refinements using diffraction data, similar to those performed on bulk materials, would be a valuable asset. Over the past decade, several advances in 3D electron diffraction (3D ED) techniques~\cite{Gemmi2019} have enabled the structure analysis of crystalline domains in thin film smaller than 100 nm~\cite{Zhang2016,Steciuk2019}. The determination of the LVO structure -- across the film's thickness -- requires to carry out 3D ED with an enhanced spatial resolution. To this end, we recently proposed to combine precession-assisted 3D ED with a line scan along the growth direction~\cite{Passuti2023}. This approach, known as scanning precession electron tomography (SPET), enables the collection of 3D ED data from multiple adjacent regions of interest (ROI) (see Methods). The first step in SPET data analysis involved the extraction of a tilt series from the SPET data stack corresponding to the DSO substrate. This provides a determination of the calibration constant (value in \AA~per pixel) and a correction of the distortions cause by the microscope's optical system and the electron beam motion~\cite{Brazda2022}. The DSO cell parameters are then used as an internal reference, enabling the lattice parameters of the LVO thin film to be determined more accurately~\cite{Passuti2023}. The diffraction patterns from regions near the surface of the film (LVO top) and near the substrate/film interface (LVO bottom) were extracted to reconstruct the precession-assisted 3D tilt series for both ROIs. During the tomography, the tilting was performed along the same direction, indicated by a black arrow in Fig.~\ref{Fig3}(a), in such a way that the probed film area was not shadowed by the substrate. The initial analysis identified an orthorhombic crystal system with a $Pbnm$ space group and an (110)-oriented LVO for both regions of the films. The refined $a$, $b$, $c$ lattice parameters for the orthorhombic cell were determined as follows: 5.5409(24)~\AA, 5.5319(21) \AA, 7.8900(18) \AA~close to the surface (LVO top) and 5.5392(21) \AA, 5.5383(21) \AA, 7.8898(18) \AA~close the film/substrate interface (LVO bottom). The unit cell volumes are 241.8 \AA$^3$ (near the surface) and 242.0 \AA$^3$ (near the interface), close to the bulk unit cell volume (242.0 \AA$^3$). $a$ and $b$ are 0.2--0.3\% higher than those determined by XRD, while $c$ remains strained to match the corresponding parameter of DSO, consistent with the LVO $[001]$ direction lying in the substrate plane. Between the two ROIS, $a$ and $b$ lattice parameters change by 0.03\% and 0.1\%, respectivley. However, this difference might be underestimated. Indeed, given the beam size used in our SPET experiment, a unit cell deformation within the first 1--3 unit cells near the interface cannot be excluded, as evidenced by the STEM-HAADF observations in Fig.~\ref{Fig1}(d--f).
\begin{table}[tp]
\caption{\label{Table3} Top: atomic positions close to the interface (LVO bottom) and the surface (LVO top). Bottom: amplitudes of the lattice distortions, given in \AA~per formula unit. }
\begin{tabular}{c|ccccc}
\diagbox{at.}{pos.} & & $x$ & $y$ & $z$ & U$_{\rm{iso}}$ (\AA$^2$) \\
\hline
 $ \rm{La}$ & & 0.4947(4) & 0.4712(3) & 0.25 & 0.0045(4)\\
 $ \rm{V}$ & LVO & 0.5 & 0 & 0 & 0.0028(6)\\
 $ \rm{O1}$ & top & 0.2179(7) & 0.2162(7) & 0.0352(7) & 0.005(1) \\
 $ \rm{O2}$ & & 0.4287(7) & -0.0097(7) & -0.25 & 0.005(2) \\
\hline
 $ \rm{La}$ & & 0.4943(3) & 0.4710(3) & 0.25 & 0.0061(5)\\
 $ \rm{V}$ & LVO & 0.5 & 0 & 0 & 0.0014(7)\\
 $ \rm{O1}$ & bottom & 0.2184(7) & 0.2188(7) & 0.0355(7) & 0.008(2) \\
 $ \rm{O2}$ & & 0.4358(7) & -0.0095(7) & -0.25 & 0.004(2) \\
\hline
\hline
 & $a^-a^-c^0$ & $a^0a^0c^+$ & X$_5^-$ & Q$_2^+$ & Q$_2^-$ \\
\hline
bulk \cite{Bordet1993}& 0.582 & 0.368 & 0.177 & 0.001 & 0.000 \\
top & 0.558 & 0.367 & 0.169 & 0.010 & 0.000 \\
bottom & 0.532 & 0.350 & 0.170 & 0.002 & 0.000\\
\end{tabular}
\end{table}

We have ultimately reached the most valuable contribution of 3D ED in general, and SPET in this study, \textit{i.e.} the quantitative crystallographic informations on the evolution of the atomic positions within the LVO film. To achieve an accurate structure refinement, we used the so-called dynamical refinement procedure to account for both dynamical scatterings effects and the precession motion of the electron beam~\cite{Palatinus2015} based on precession-assisted 3D ED data (see Methods and Supplementary Information). Comparing the atomic positions reported in Table~\ref{Table3}, the top and bottom LVO regions exhibit the same structure within standard deviations. The refinement also indicates no atomic vacancies, as suggested by the low values of the atomic displacement parameters U$_{\rm{iso}}$. These results are consistent with the findings from STEM-HAADF observations, which uses a smaller probe size and smaller scanning step size. In particular, from the full structure refinements, the displacement of the La$^{3+}$ cations projected onto the $[\bar{1}10]$ direction was determined to be 22.7~pm, in excellent agreement with the displacements obtained from the STEM-HAADF image analysis. We have also performed a symmetry mode analysis of the structures. The results are summarized in Table~\ref{Table3} for the experimental structures and as for the DFT calculated structures, the distortion modes are the $a^-a^-c^0$ and $a^0a^0c^+$ oxygen octahedral rotation modes, the X$_5^-$ mode and the Q$_2^+$ and Q$_2^-$ cooperative JT modes. The amplitude of the distortion modes are reduced as we move away from the 0 K calculated phases (see Table~\ref{Table1}) and different from the bulk values. As for the bulk material, we observe that only the Q$_2^+$ mode is active at room temperature, reminiscent of a $Pbnm$ space group ($P2_1/m$ in thin film). The oxygen octahedral rotation and JT amplitudes increase across the film thickness towards values close to the bulk materials (within our experimental accuracy). We have determined the oxygen octahedral rotation angles [see Fig.~\ref{Fig3}(b)] which are reduced compared to the bulk materials ($\alpha=\beta=8.800^{\circ}$ and $\gamma=7.555^{\circ}$~\cite{Bordet1993}). While the Q$_2^+$ remains quenched across the thickness at a value lower than the bulk, the amplitude of Q$_2^+$ is 10 times higher at the surface compared to the interface and the bulk. The DFT calculations showed that G$\rm_{SO}$ and C$\rm_{SO}$ states are close in energy. Hence, the Q$_2^+$/Q$_2^-$ imbalance between the top and the bottom of the film may favor the coexistence and/or a gradient of G$\rm_{SO}$/C$\rm_{SO}$ ordered spin populations (and the associated C$\rm_{OO}$/G$\rm_{OO}$) throughout the film thickness below the N\'eel temperature~\cite{Miyasaka2003}. For example, the N\'eel vector associated to G$\rm_{SO}$ and C$\rm_{SO}$ may be rotated by using external stimuli such as strain fields in $(110)$-oriented RVO$_3$ heterostructures or susceptible to be manipulated by an electric or magnetic field.

%
%
\section{Conclusion}
In summary, we have shown an in-depth structural investigation of a 50~nm-LaVO$_3$ thin film, grown onto (110)-oriented DyScO$_3$. The theoretical and experimental investigations show that LaVO$_3$ adopts the bulk-orthorhombic $Pbnm$ space group and is epitaxially strained by the substrate, which imposes its (110) crystallographic orientation to the 140 deposited unit cells. We have also reached quantitative crystallographic information on the atomic positions within the perovskite thin film. This allowed to perform a symmetry mode analysis to determine the relevant structural distortion modes of the grown heterostructure. Because the spin-orbital interactions are strongly correlated to structural distortions, the experimental determination of reliable atomic positions in thin films is an asset for modeling the properties of vanadates and other transition-metal perovskites.
\section{METHODS}
\textbf{First-principle calculations}. The first principles calculations were performed using density functional theory (DFT) with the VASP package~\cite{Kresse1993,Kresse1996} and employing the PBEsol functional. In addition, we considred a U potential on V $3d$ levels of 3.5 eV, entering as a single effective parameter~\cite{Duradev1998}. This parameter was fitted in Ref.~\cite{Varignon2015}, providing, in this study, correct electronic, magnetic, and structural features for the LaVO$_3$ ground state. We used the projected augmented wave (PAW) potential to model the core electrons~\cite{Blochl1994}. In the calculations, a ($2a$, $2a$, $2a$) cubic cell was considered, which allows for the oxygen octahedra rotations and Jahn-Teller motions to develop. The energy cutoff is set to 500 eV and a $4\times4\times4$ $k$-point mesh is employed.

\textbf{Thin film growth}. The (110)-oriented DSO substrates, provided by Crystec GmbH ($a=5.44$~\AA, $b=5.71$~\AA~and $c=7.89$~\AA), were introduced in a MBE chamber (DCA Instruments) with a base pressure of 6$\times 10^{-11}$~mbar, equipped with reflection high-energy electron diffraction (RHEED). The atoms are evaporated using high temperature Knudsen cells, whose atomic fluxes are calibrated using a quartz crystal microbalance before and after the deposition. The Pr, La and V atomic fluxes were fixed to $1.5 \times 10^{13}$~atom.cm$^{-2}$.s$^{-1}$ with 5\% accuracy. The growth conditions have been optimized to obtain high quality single phase films. To get the proper oxygen stoichiometry, an oxidant gas (O$_3$+O$_2$) was introduced at a pressure fixed in the 5-7$\times 10^{-7}$~mbar range in the chamber. During the growth, the substrate temperature was kept at 850$^{\circ}$C.

\textbf{X-ray diffraction}. The crystallinity and the structure were characterized by X-ray diffraction (XRD) with a PANalitycal X'pert Pro MRD diffractometer using monochromatic Cu K$\alpha_1$ radiation ($\lambda=1.54056$~\AA). For high resolution reciprocal space mapping, the XRD measurements were performed in skew geometry using a triple axis crystal analyzer.  

\textbf{Transmission electron microscopy}. The atomically resolved microstructure was determined by cross-sectional high-resolution transmission electron microscopy (HRTEM) and high angle annular dark field (HAADF) scanning transmission electron microscopy (STEM). The cross-section lamella were prepared by focus ion beam with Ga$^+$ ion milling (FEI-Helios Nanolab 600i) and observed with a double-aberration corrected JEOL ARM microscope operated at 200~kV. 

\textbf{Scanning precession electron diffraction tomography}. The electron diffraction analysis was performed on a cross-section lamella using a JEOL F200 cFEG microscope working at 200 kV. Scanning Precession Electron diffraction Tomography (SPET) was carried out on the sample by scanning the electron beam (15~nm in diameter) in a direction perpendicular to the film/substrate interface. Every line scan was performed using the NanoMEGAS DigiSTAR unit on a 250~nm line going from the polycrystalline coating to the substrate with a step of 2.5~nm. The acquisitions were carried out in a tilt range of 75$^{\circ}$ with a tilt step of 1$^{\circ}$ and a precession-semi angle of 1.2$^{\circ}$. The diffraction patterns were acquired with an ASI Cheetah M3 512$\times$512 CMOS hybrid pixel detector using the ACCOS GUI from ASI, each with an exposure time of 0.5~second. In view of this description, a SPET experiment actually corresponds to the acquisition of so-called precession-assisted 3D ED datasets on multiple regions of interest located along a line.  Each 3D ED datasets was processed using PETS2~\cite{Palatinus2019} for indexation, cell parameters refinement and intensity integration. Subsequent structure refinements were carried out using the JANA2020 program~\cite{Petricek2023}.
\begin{acknowledgement}
This work is supported by the Région Grand Est projects RHUM (AAP-013-075) and DECOX (23JC-004), the France 2030 government investment plan under the ANR grant PEPR SPIN - SPINMAT (ANR-22-EXSP-0007), and the ANR project CITRON (ANR-21-CE09-0032). The work has benefited of the resources of experimental platforms at IJL: the Tube Davm, 3M (funded by FEDER EU, ANR, Région Grand Est, and Métropole Grand Nancy), and XGamma, all supported by the LUE-N4S project, part of the French PIA project Lorraine Université d’Excellence (ANR-15IDEX-04-LUE), and by FEDER-FSE Lorraine et Massif Vosges 2014-2020, an EU program. The resources for the first-principles calculations were provided by the High Performance Computing resources of CRIANN through the projects 2020005 and 2007013 and of CINES through the DARI project A0080911453. Part of this work was done in the framework of NanED project funded by the European Union (ITN MSCA grant agreement no. 956099).  
\end{acknowledgement}


\end{document}